\documentclass{moriond}

\def\be{\begin{equation}}
\def\ee{\end{equation}}
\def\bea{\begin{eqnarray}}
\def\eea{\end{eqnarray}}

\newcommand{\Photo}{\includegraphics[height=35mm]{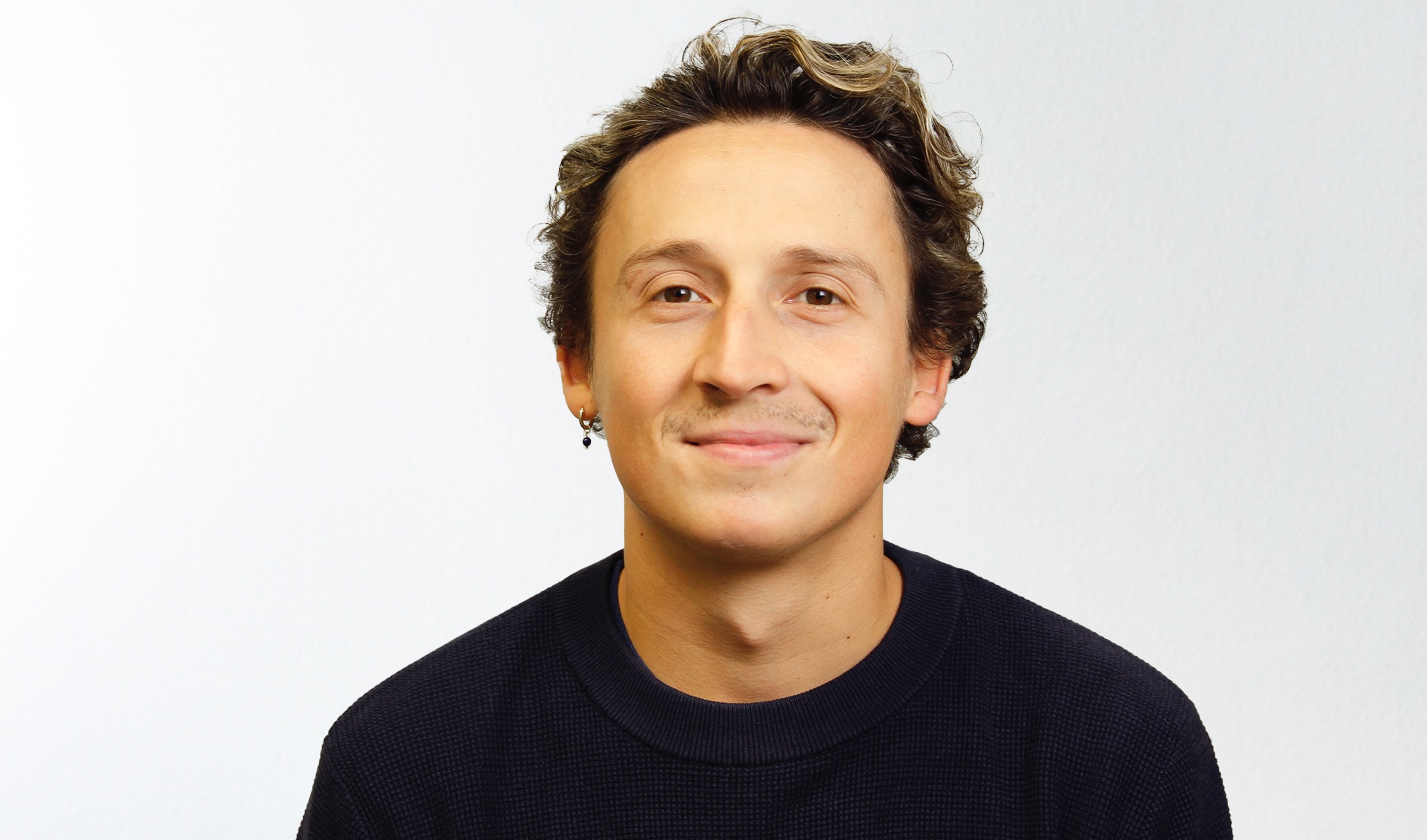}}

\usepackage{amsmath}

\begin{document}
\vspace*{4cm}
\title{Latest results from the IceCube Neutrino Observatory}

\author{ T.J. van Eeden\\On behalf of the IceCube Collaboration }

\address{
DESY,\\
Platanenallee 6, 15738 Zeuthen
}

\maketitle\abstracts{
The IceCube Neutrino Observatory has opened a new window into the high-energy Universe, providing measurements of neutrinos over a broad energy range. This contribution presents recent results, including a follow-up on the first identification of a steady neutrino source NGC 1068, measurements of the flavor composition of the diffuse astrophysical flux, limits on prompt atmospheric neutrinos, and searches for neutrinos from dark matter annihilation in the Sun. These measurements probe neutrino production mechanisms, fundamental particle interactions, and physics beyond the Standard Model. Looking forward, the recently deployed IceCube Upgrade will enhance sensitivity to lower-energy neutrinos and reduce systematic uncertainties, while the planned IceCube-Gen2 will expand the detector volume, increase the neutrino detection rate, and extend energy reach, enabling more detailed studies of cosmic sources and high-energy particle physics.
}

\section{Introduction}

Neutrinos are electrically neutral and weakly interacting fermions that provide a probe of fundamental physics over a wide range of energies. The observation of neutrino oscillations demonstrates that neutrinos have nonzero masses and mix between flavors, which establishes physics beyond the Standard Model and kicked off precision studies of mixing parameters and matter effects. Atmospheric neutrinos, produced in cosmic ray interactions in the atmosphere, provide controlled baselines for oscillation measurements and sensitivity to Earth matter effects. Astrophysical neutrinos extend the accessible energy range by several orders of magnitude and probe particle acceleration in extreme environments while enabling measurements of neutrino cross sections at energies beyond the reach of accelerators. These neutrinos are being observed by the IceCube Neutrino Observatory and this contribution highlights several recent results on atmospheric and astrophysical neutrino measurements, including implications for particle physics and astroparticle studies.

\section{The IceCube detector}

The IceCube Neutrino Observatory is a cubic-kilometer-scale Cherenkov detector installed in the glacial ice at the geographic South Pole \cite{icecube2017detector}. It consists of 86 vertical strings instrumented with 5160 optical modules deployed at depths between about 1.5 km and 2.5 km, forming a three-dimensional array that records the arrival time and amplitude of photons in the ice. The detector enables the reconstruction of neutrino-induced events over a broad energy range, from a few GeV to beyond the PeV scale. A denser sub-array, IceCube DeepCore, is located in the clearest ice near the bottom center of the detector and lowers the effective energy threshold to the order of 10 GeV, which is essential for precision studies of atmospheric neutrinos and oscillation effects \cite{deepcore2012design}. Neutrinos are detected indirectly via charged particles produced in neutrino interactions with nuclei in the ice or the surrounding bedrock. These secondary particles emit Cherenkov radiation when traveling faster than the speed of light in ice, and the resulting light pattern is used to infer the energy, direction, and topology of the original neutrino. 

\section{Latest results from IceCube}

\subsection{Neutrino Emission from X-ray-bright Active Galactic Nuclei}

\begin{figure}[h!]
    \centering
    \includegraphics[width=\linewidth]{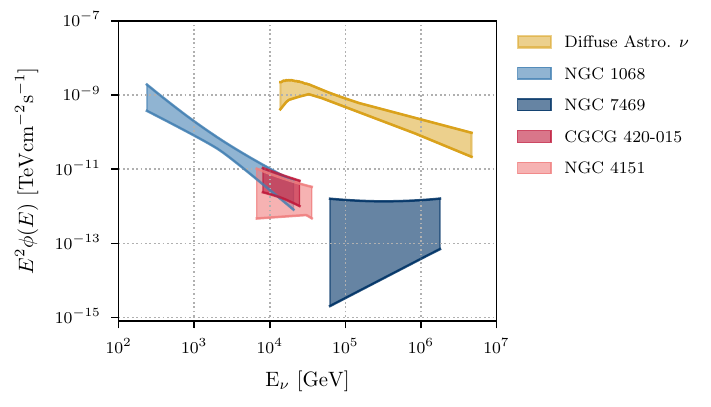}
    \caption{All-flavor neutrino flux of the top 4 X-ray-bright AGNs \protect\cite{seyfertfollowup2026}. The shown fluxes correspond to best-fit estimates and do not represent statistically significant individual detections. These results are compared with the diffuse astrophysical neutrino flux \protect\cite{break2026}. Figure adapted from Ref. \protect\cite{seyfertfollowup2026}.}
    \label{fig:xray}
\end{figure}

The first compelling evidence for a steady astrophysical neutrino source was reported by the IceCube Neutrino Observatory in 2022 with the detection of a significant excess of neutrinos from the nearby active galaxy NGC 1068 \cite{ngc2022}. Interestingly, the source shows unexpectedly weak gamma-ray emission, while being one of the brightest X-ray active galactic nucleus (AGN) in the sky. Based on this result, a recent analysis uses 13.1 years of data to perform a targeted search for neutrino emission from X-ray-bright AGNs in the northern sky and confirms NGC 1068 as the most significant steady emitter \cite{seyfertfollowup2026}. The neutrino spectrum of the source is characterized by a soft unbroken power-law with a spectral index of $\gamma = 3.4\pm0.2$ and is shown in Figure \ref{fig:xray}. 

\begin{figure}[h!]
    \centering
    \includegraphics[width=0.8\linewidth]{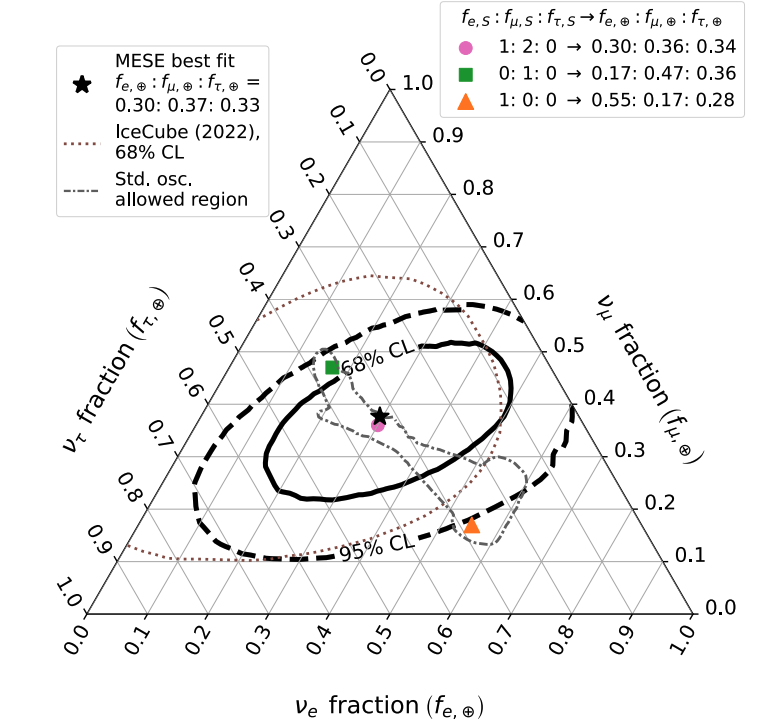}
    \caption{Results of the flavor composition measurement together with expected compositions at Earth for different neutrino production mechanisms \protect\cite{flavor2025}. The benchmark scenarios correspond to pion decay ($\nu_e:\nu_\mu:\nu_\tau = 1:2:0$ at the source), muon damped ($0:1:0$), and neutron beam ($1:0:0$), after propagation and flavor mixing to Earth. Figure adapted from Ref. \protect\cite{flavor2025}.}
    \label{fig:flavor}
\end{figure}

The observed neutrino emission is likely generated in the immediate vicinity of the supermassive black hole that powers the AGN, with the most plausible site being the corona \cite{murase2024encyclopedia}. This region consists of a plasma of hot electrons with temperatures of the order $10^{9} \text{ K}$ and is responsible for the characteristic X-ray emission of AGNs. In this environment, ultraviolet photons are Compton up-scattered by energetic electrons to keV energies, producing the observed X-ray spectrum. These X-ray photons can act as targets for photomeson production when interacting with $\sim 100 \text{ TeV}$ protons leading to neutrino production in the observed energy range. The study further investigates a sample of 47 X-ray-bright Seyfert galaxies and finds an excess at the level of 3.3$\sigma$ from an ensemble of 11 sources, providing evidence for a population of neutrino sources associated with AGN cores and supporting scenarios in which high-energy neutrinos are produced in dense environments near supermassive black holes. The neutrino spectra of the top 4 sources are shown in Figure \ref{fig:xray}, together with the most recent measurement of the diffuse astrophysical neutrino flux \cite{break2026}.

\subsection{Characterization of the flavor composition of astrophysical neutrinos}

The flavor composition of astrophysical neutrinos provides a direct probe of the production mechanisms at their sources and of their propagation over cosmic distances \cite{arguelles2015effect}. Neutrinos are typically produced in hadronic interactions with an initial flavor ratio close to $\nu_e:\nu_\mu:\nu_\tau = 1:2:0$ via the decay of charged pions and kaons. The effect of neutrino oscillations over astrophysical baselines leads to an approximately equal flavor composition at Earth. Deviations from this expectation can indicate different neutrino production mechanisms or the presence of new physics affecting neutrino propagation.

Recent work showed the measurement of the flavor composition of the diffuse astrophysical neutrino flux using 11.4 years of data from the IceCube \cite{flavor2025}. This analysis uses a sample of events that have their interaction vertices contained inside the detector volume. These 'starting events' are sensitive to all neutrino flavors using track, cascade and double cascade classifications. Track-like events are primarily sensitive to the $\nu_\mu$ component, cascade events to $\nu_e$, and double-cascade topologies to $\nu_\tau$. The best-fit flavor ratio at Earth is found to be approximately $f_e:f_{\mu}:f_{\tau} = 0.30:0.37:0.33$ and is shown in Figure \ref{fig:flavor}. This is consistent with expectations from standard three-flavor neutrino oscillations over cosmic baselines. These results provide important constraints on the production mechanisms at the sources and disfavor extreme scenarios such as pure neutron decay at the 99\% confidence level.

\subsection{Prompt atmospheric neutrinos}

The IceCube Neutrino Observatory has established a diffuse flux of high-energy astrophysical neutrinos \cite{icecube2013evidence,break2026}, but precision studies of this flux require careful treatment of backgrounds. A sub-dominant component is the prompt atmospheric neutrino flux, arising from the decay of charmed mesons produced in cosmic ray air showers. The production of charm in the very forward region of hadronic interactions remains poorly constrained, and its study provides a natural link to accelerator-based measurements of charm production and nucleon structure at high energies. Constraining the prompt flux is therefore not only essential for reducing systematic uncertainties in astrophysical neutrino analyses but also informs our understanding of processes in kinematic regimes inaccessible to current accelerators. 

A recent study from IceCube combines cascade-like and track-like event samples to maximize sensitivity to the prompt component, accounting for uncertainties in conventional atmospheric and astrophysical fluxes \cite{prompt2025}. The resulting best-fit flux is consistent with zero, yielding an upper limit of $4\times10^{-16} \text{ GeV}^{-1} \text{ s}^{-1} \text{ m}^{-2} \text{ sr}^{-1}$ at 10 TeV and providing constraints on charm production in the atmosphere and its connection to high-energy hadronic physics studied in accelerators.

\begin{figure}[h!]
    \centering
    \includegraphics[width=0.85\linewidth]{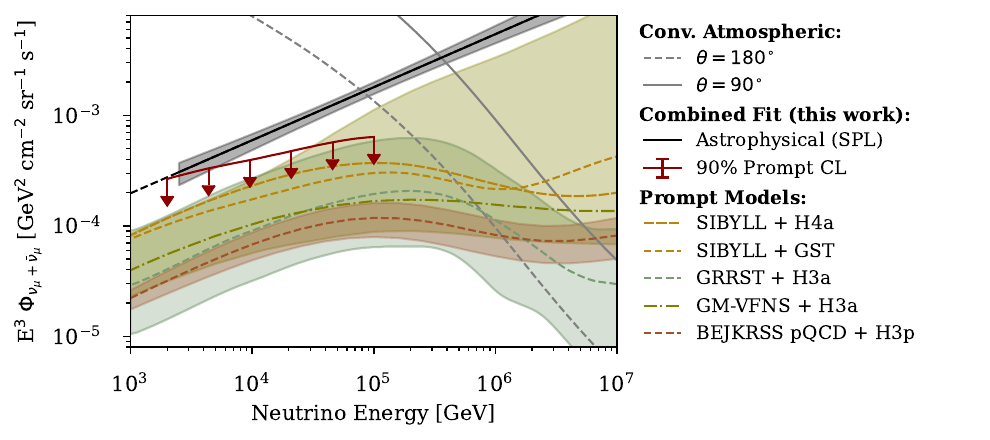}
    \caption{The upper limit on the prompt neutrino flux and the best-fit single power law (SPL) spectrum of astrophysical neutrinos \protect\cite{prompt2025}. Predictions for the prompt neutrino flux are included for several models. The conventional atmospheric neutrino flux (Conv.) is shown for vertical ($\theta = 180^\circ$) and horizontal ($\theta = 90^\circ$) directions. Figure adapted from Ref. \protect\cite{prompt2025}.}
    \label{fig:flavor}
\end{figure}

\subsection{Dark matter in the Sun}

Dark matter is a component of the Universe that dominates the matter content and is inferred from its gravitational effects on galaxies and cosmology. A leading candidate is the class of weakly interacting massive particles (WIMPs). If these WIMPs annihilate or decay, they can produce Standard Model particles, including high-energy neutrinos, which can escape dense environments and provide a detectable signal for indirect dark matter searches.

A recent study targets dark matter in the Sun using 9.3 and 10.4 years from the DeepCore and IceCube detectors \cite{sun2025}. The WIMPs can scatter off nuclei in the Sun, lose energy, and become gravitationally trapped. Over time, repeated scatterings lead to an accumulation of dark matter in the solar core. Annihilations of these particles produce Standard Model products, but only neutrinos can escape the dense solar interior. Consequently, an excess of high-energy neutrinos from the Sun provides a direct avenue for indirect dark matter searches. Because capture and annihilation are expected to reach equilibrium in the Sun, these searches are particularly sensitive to the spin-dependent dark matter–proton scattering cross section. The constraints of the scattering cross section are world-leading above dark matter masses of 100 GeV, and are shown in Figure \ref{fig:flavor}. The results are compared to other direct and indirect detection methods. 

\begin{figure}[h!]
    \centering
    \includegraphics[width=0.6\linewidth]{figures/sun.png}
    \caption{Limits of IceCube on the spin-dependent dark matter-proton scattering cross section compared with limits from direct and indirect detection methods. Figure adapted from Ref. \protect\cite{sun2025}.}
    \label{fig:flavor}
\end{figure}

\section{Outlook and conclusion}

The recent IceCube results presented here demonstrate significant progress in neutrino astronomy. The first identification of a steady neutrino source from NGC 1068, measurements of the diffuse astrophysical flavor composition, and limits on prompt atmospheric and solar dark matter induced neutrinos have collectively improved our understanding of high-energy neutrino sources, particle interactions, and potential physics beyond the Standard Model.

The IceCube Upgrade, deployed at the South Pole during the 2025–2026 austral summer season, adds densely instrumented strings with enhanced optical modules and dedicated calibration devices \cite{upgrade2019}. This upgrade will enhance sensitivity to lower-energy neutrinos, improving measurements of neutrino oscillation parameters, searches for dark matter, among other topics. The new calibration devices will allow more detailed studies of systematic uncertainties, particularly in the ice, improving the quality of both future and previously recorded data.

The planned IceCube‑Gen2 will enhance the existing IceCube detector at the South Pole, expanding the instrumented volume by a factor of eight and incorporating complementary arrays for radio, and surface detection \cite{gen2021}. This next generation instrument is designed to resolve the high-energy neutrino sky from TeV to EeV energies, investigate cosmic particle acceleration through multimessenger observations, reveal the sources and propagation of the highest-energy particles, and probe fundamental physics. IceCube‑Gen2 will increase the annual detection rate of cosmic neutrinos by roughly an order of magnitude, detect sources up to five times fainter than previously possible, and extend the energy reach by several orders of magnitude with the addition of a large-scale radio array.

\section*{References}
\bibliography{thijsvaneeden}

%%% manually generated bibliography
%\begin{thebibliography}{99}
%\bibitem{ja}C Jarlskog in {\em CP Violation}, ed. C Jarlskog
%(World Scientific, Singapore, 1988).
%\bibitem{ma}L. Maiani, \Journal{\PLB}{62}{183}{1976}.
%\bibitem{bu}J.D. Bjorken and I. Dunietz, \Journal{\PRD}{36}{2109}{1987}.
%\bibitem{bd}C.D. Buchanan {\it et al}, \Journal{\PRD}{45}{4088}{1992}.
%\end{thebibliography}

\end{document}